\documentclass[12pt]{iopart}

\usepackage{epsfig}
\usepackage{iopams}
\newcommand{\beq} {\begin{equation}}
\newcommand{\eeq} {\end{equation}}

\newcommand{\pon} {p_1}
\newcommand{\ptw} {p_2}
\newcommand{\hon} {h_1}
\newcommand{\htw} {h_2}
\newcommand{\bfsigma}{\mbox{\boldmath $\sigma$}}
\newcommand{\bfp}{{\bf p}}
\newcommand{\bqu}{{\bf q}}

\begin{document}
\vspace*{-2cm}

\centerline{\large {\bf Photo-emission of two protons from nuclei}}

\vspace{.3cm}

\begin{center}
{\bf Marta Anguiano$^1$, Giampaolo Co'$^{2,3}$ and Antonio M. Lallena$^4$} \\

\vspace{.1cm}
{\small 
$^1$ Departamento de Radiaci\'on Electromagn\'etica,
Instituto de F\'{\i}sica Aplicada, \\CSIC, Serrano 144, E-28006 MADRID,
Spain \\
$^2$ Dipartimento di Fisica, Universit\`a di Lecce, I-73100
Lecce, Italy \\
$^3$ Istituto Nazionale di Fisica Nucleare  sez. di Lecce, I-73100
Lecce, Italy \\
$^4$ Departamento de F\'{\i}sica Moderna, Universidad de
Granada, \\E-18071 Granada, Spain
}
\end{center}

\vspace{.2cm}

\small{
The photo-emission of two protons from the $^{12}$C, $^{16}$O and
  $^{40}$Ca nuclei is investigated. Aim of the work is the study of
  the possibilities offered by this probe to obtain information about
  the characteristics of the short-range correlations.  We have also
  evaluated the effects of the two-body $\Delta$-currents which, in
  this processes, compete with those produced by the short-range
  correlations. Our results show that ($\gamma$,pp) processes could
  be more useful than (e,e'pp) for the study of the short-range
  correlations. 
} 

\vspace{.3cm}

\section{Introduction}
\label{sec:intro}
This article belongs to a series dedicated to the study of the Short
Range Correlations (SRC) effects in processes induced by
electromagnetic probes on atomic nuclei. We have already conducted
this study for inclusive electron scattering \cite{co98,mok00,co01},
(e,e'p) \cite{mok01}, ($\gamma$,p) \cite{ang02} and (e,e'pp)
\cite{ang03} reactions.  In this work we investigate the ($\gamma$,pp)
case.

Inclusive and one-nucleon emission processes are dominated by the
uncorrelated one-body responses. Our results
\cite{co98,mok00,co01,mok01,ang02} show that the SRC produce small
corrections to the uncorrelated cross sections. The size of these
corrections remains within the theoretical uncertainties due to the
arbitrary choice of the input parameters.

The emission of two nucleons cannot be induced by uncorrelated
one-body operators.  For this reason these processes are the ideal
tool to study the SRC effects. However, even in this case the
situation is not completely clean, since two nucleons can also be
emitted by two-body Meson-Exchange Currents (MEC) without any
contribution of the SRC. The most important MEC terms, called seagull
and pionic, are produced by the exchange of charged pions.  When the
two emitted nucleons are of the same type, for example two protons,
these terms do not act. For this reason the MEC contribution in the
emission of a proton-neutron pair is noticeably larger than in the
case of nucleon-like emission \cite{giu98,ryc98}. In this last case
the active MEC terms, where a single uncharged meson is exchanged, are
due to the virtual excitations of the $\Delta$ resonance
\cite{ama94}.

In principle the contribution of the MEC could be eliminated by a
super Rosenbluth separation of the (e,e'pp) cross section.  With this
procedure it is possible to extract the longitudinal (charge)
response, where the MEC effects are negligible \cite{lal97}, therefore
the signature of the SRC would be very clean.

From the experimental point of view the Rosenbluth separation is
extremely difficult.  We think it is more fruitful to study the effects of
the the $\Delta$ currents in order to keep them under control. In
\cite{ang03} this investigation has been done for (e,e'pp) processes,
and in the present paper we continue it for the case of real photons.
Real photons have less flexibility than electrons, because the
momentum transferred to the nucleus is fixed by the energy of the
absorbed photons. On the other hand, since they are purely transverse,
the longitudinal terms are not present in the cross section, and this
simplifies the analysis of the results.

In recent years theoretical studies of the emission of two nucleons
induced by the absorption of real photons have been done by the
Valencia \cite{car92}, Pavia \cite{giu92,sch04} and Gent \cite{van94}
groups. The Valencia model is based on a Fermi gas description of the
nucleus, i.e. all the single particle wave functions are plane
waves. The final state interactions (FSI) are calculated by explicitly
considering the various diagrams that describe the interaction of the
emitted nucleons with the rest nucleus. Our approach is more similar
to that of the Pavia group which uses Woods-Saxon single particle wave
functions and complex optical potentials to treat FSI. The difference
in the two models is in the treatment of the SRC. The Pavia group uses
two-nucleon spectral functions, while we make an expansion of the
transition matrix elements in power of the correlation function (see
section \ref{sec:cs}). The Gent group uses single particle wave
functions generated by Hartree-Fock calculations. FSI are considered
within a self-consistent continuum Random Phase Approximation
framework.

The experimental situation is very promising. Data have been taken at
various laboratories \cite{kan87} and have shown the feasibility of
these experiments with the present facilities.  Furthermore, they have
clearly indicated the different sensitivities of ($\gamma$,pn) and
($\gamma$,pp) reactions to the SRC and MEC.  A direct comparison
between our results and the available data is not straightforward,
since the data have been averaged on the energy and angular
acceptances of the nucleon detectors; also a certain range of the
energy of the incident photons has been considered. For this reason
we postpone this comparison, requiring a direct contact with the
experimental groups, to eventual forthcoming publications.

In this paper after
summarizing in section \ref{sec:cs} the basic formulas used
to describe ($\gamma$,pp) processes with our model,
we discuss the results obtained for the $^{12}$C,
$^{16}$O and $^{40}$Ca nuclei aimed to understand our capacity of
keeping under control the $\Delta$ currents and to investigate the SRC
characteristics.

\section{The cross section}
\label{sec:cs}
In this section we briefly recall the expressions used in our
calculations to describe the ($\gamma$,pp) cross section.  A more
detailed derivation of the cross section expression can be found in
\cite{bof96}, while our nuclear model has been widely described in
\cite{ang03}.

\begin{figure}[b]
\begin{center}
\includegraphics*[width=7cm]{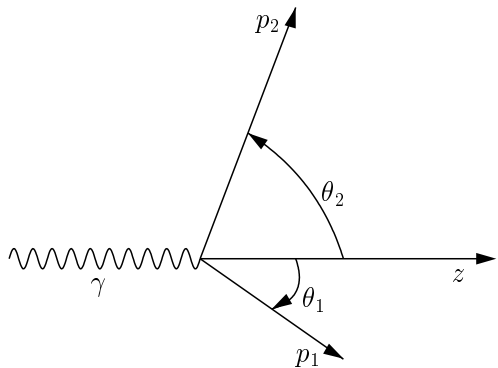}
\end{center}
\caption{\small Reference system used in our calculations. The $z$
  axis indicates the direction of $\bqu$, the photon momentum.  The
  vectors $\bqu$, ${\bf p}_1$ and ${\bf p}_2$ are on the same
  plane. The arrows of the angles $\theta_1$ and $\theta_2$ indicates
  the direction of their positive values.}
\label{fig:axis}
\end{figure}

We label the energy and momentum of the photon as $\omega$ and $\bqu$,
respectively, and the four momenta of the emitted nucleons with $\pon
\equiv (\epsilon_1,\bfp_1)$ and $\ptw \equiv (\epsilon_2,\bfp_2)$.
The kinematics set up is defined in Fig. \ref{fig:axis}. The
directions of the two emitted nucleons can be identified by the angles
formed with the photon direction.  We call these angles $\theta_1$ and
$\theta_2$, and define $\theta_{12}=\theta_1-\theta_2$. The figure
shows that, in our representation, the two angles are defined positive
in opposite directions with respect to $z$-axis.

In natural units ($\hbar=c=1, e^2=1/137.04$), the 
expression of the cross section we adopt is:
\label{sec:xsect}
\beq
\frac{{\rm d}^5 \sigma}
{ {\rm d}\Omega_1
{\rm d}\epsilon_2 {\rm d}\Omega_2 } \, = \, 
\frac{K}{(2 \pi)^6} \, 
\frac{2 \pi^2 e^2}{\omega}
\,  f_{{\rm rec}} \, w_t(|\bqu|) \, ,
\label{eq:cross}
\eeq
where $e$ is the elementary charge.
In the above equation, 
$K$ is a kinematic factor
\beq
K\, = \, m^2 \, |\bfp_1| \,\, |\bfp_2| \, ,
\eeq
and $f_{{\rm rec}}$ is a recoil factor defined as:
\beq
\label{eq:rec}
f_{{\rm rec}}^{-1}\, =\,  1\, +\,  \frac{m}{M_{A-2}} \, 
\left[ 1+ \frac{|\bfp_1|}{|\bfp_2|}\cos \theta_{12} - 
          \frac{|\bqu|}{|\bfp_2|}\cos \theta_2  \right] \, ,
\eeq
where $m$ is the proton mass and $M_{A-2}$ the mass of the rest
nucleus.

The nuclear structure information is included in the $w_t$
factor
\begin{eqnarray}
\label{eq:wt}  
w_t(|\bqu|) & = & 
\langle \Psi_i | J_{-}^\dagger (\bqu) | \Psi_f \rangle 
      \,   \langle \Psi_f | J_{-} (\bqu) |\Psi_i \rangle \\
&& \nonumber + \, 
        \langle \Psi_i | J_{+}^\dagger (\bqu) | \Psi_f \rangle 
      \,   \langle \Psi_f | J_{+} (\bqu) |\Psi_i \rangle \, , 
\end{eqnarray}
where we have indicated with $|\Psi_i\rangle$ and $|\Psi_f\rangle$
the initial and final states of the full hadronic system.  We do not
consider the polarization of the emitted nucleons, therefore in the
previous equations a sum on the third components of the spin of the
emitted particles and of the angular momentum of the residual
nucleus is understood.

The operators present in Eq. (\ref{eq:wt}) are the transverse
components of the electromagnetic four-current operator composed by
one- and two-body terms. The one-body electromagnetic currents 
are produced by the convection current,
\begin{equation}
\label{eq:conv}
J^{\rm C}({\bf r}) \, = \, 
\sum^A_{k=1} \frac{-i}{2m_k} \,\frac{1+\tau^3_k}{2} 
\left[\delta({\bf r}-{\bf r}_k) \nabla_k - 
      \nabla_k \delta({\bf r}-{\bf r}_k) \right] \, ,
\end{equation}
and by the magnetization current,
\begin{equation}
\label{eq:mag}
J^{\rm M}({\bf r}) \, = \, 
\sum^A_{k=1} \frac{1}{2m_k} \,
\left(\mu^{\rm P}\frac{1+\tau^3_k}{2} + 
\mu^{\rm N}\frac{1-\tau^3_k}{2} \right)
\nabla \times \delta({\bf r}-{\bf r}_k) \, \bfsigma_k \, .
\end{equation}
In the previous equations $m_k$ indicates the mass of $k$--th nucleon,
$\mu^{\rm P}$ and $\mu^{\rm N}$ the anomalous magnetic moment of the
proton and of the neutron respectively, $\bfsigma_k$ the Pauli spin matrix
of the $k$-th nucleon and $\tau^3_k=1$ ($-1$) in case the $k$-th
nucleon is a proton (neutron). In our calculations the internal
structure of the nucleon has been considered by folding the point-like
responses with the electromagnetic nucleon form factors of
\cite{hoe76}.

We point out that, in the present article, we have considered the
convection current which was not included in the calculation of
\cite{ang03}. We have shown in \cite{ang02} that the contribution of
the convection current is relevant for the photoabsorption processes,
while it is negligible for electron scattering in the quasi-elastic
region.

\begin{figure}
\begin{center}
\includegraphics*[width=7cm]{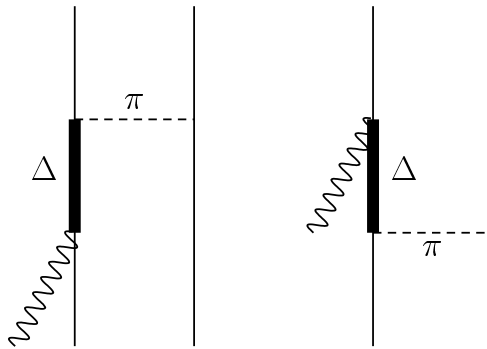}
\end{center}
\caption{\small Meson exchange current diagrams considered in our
       calculations. The wavy lines represent the photon, and the full
       lines the nucleons. Since we describe the emission of two
       protons, the exchanged pion is chargeless.}
\label{fig:feydelta}
\end{figure}

We have already mentioned in the introduction that,
since we have considered only the emission of two like particles,
specifically two protons, the two-body currents induced by the
exchange of charged mesons do not contribute. The main two-body
current contribution arises from diagrams like those of Fig.
\ref{fig:feydelta} where the exchanged $\pi$ meson is chargeless.  We
describe these $\Delta$ currents following the prescription of
\cite{wil96}. The corresponding expressions in our approach are given
in \cite{ang03}. We would like to mention here that these currents
depend on the product of the three coupling constants related to the
three vertexes of the diagrams of Fig. \ref{fig:feydelta}. They are:
the pion-nucleon coupling constant $f_{\pi N N}$, the
pion-nucleon-delta coupling constant $f_{\pi N \Delta }$, and,
finally, the photon-nucleon-delta coupling constant $f_{\gamma N
  \Delta}$. Out of these three constants only $f_{\pi N N}$ is rather
well determined by the experimental data.  Its commonly accepted value is
$f^2_{\pi N N}/4 \pi$=0.079. We used the uncertainty on the values of
the other constants to estimate the sensitivity of our results to the
two-body $\Delta$ currents.

The calculation of the nuclear response (\ref{eq:wt}) requires a model
for nuclear states. A detailed description of our model can be found
in \cite{ang03}, and here we recall some of the basic ideas and
approximations.

The nuclear states are described as:
\beq
| \Psi \rangle \, =  \, 
\frac { F |\Phi \rangle }
{ \langle \Phi | F^+  F | \Phi \rangle } \, .
\label{eq:nstate}
\eeq In the above expression $|\Phi \rangle$ is a Slater determinant
formed by single particle states generated by a one-body hamiltonian,
which in our calculations is obtained by using a Woods-Saxon potential
well.  The ground state is described by using the ground state Slater
determinant, formed by all, and only, the states whose single particle
energies are smaller than the Fermi energy. The excited states are
described as Slater determinants formed by two-particle two-hole
excitations. The two particle states are continuum single particle
wave functions, calculated as scattering waves from the mean-field
potential.

The ingredient making possible the two-nucleon emission  by
one-body operators, is the correlation function $F$.
In our calculations we have used purely scalar correlations defined
as:
\beq
F(1,2,...A) \, = \, \prod^A_{i<j} f(r_{ij}) \, ,
\label{eq:jastr}
\eeq 
where $r_{ij}=|{\bf r}_i-{\bf r}_j|$ is the distance between 
the particles $i$ and $j$. The insertion of Eq.
(\ref{eq:nstate}) in the expression (\ref{eq:wt}) allows us to make an
expansion in terms of the function $h_{ij}= f^2(r_{ij})-1$.  The main
approximation of our model consists, in retaining only those terms of
the expansion where the function $h_{ij}$ appears only once.  

At this point, it may be worth to mention two properties related to
the presence of the denominator in Eq. (\ref{eq:nstate}). The first,
and more relevant one from the physics point of view, is that the
denominator ensures the proper normalization of the many-body
correlated wave function.  The second property is more technical, but
still relevant. The denominator cancels all the unlinked terms of the
expansion in terms of $h_{ij}$ \cite{fan87}. Only after making this
cancellation we cut the expansion to the first order in $h_{ij}$.

This model has been developed to calculate the ground state densities 
and momentum distributions \cite{co95,ari97}, and later extended to
describe the nuclear excited states
\cite{co98,mok00,co01,mok01,ang02,ang03}.  The validity of the truncation
procedure has been tested in nuclear matter against a calculation
where all the power terms in $h_{ij}$ have been considered
\cite{ama98}. The quasi-elastic charge responses obtained with the two
methods are in excellent agreement.

\section{Results}
\label{sec:res}
The kinematics of the process is such that
energy and momentum conservations imply:
\begin{eqnarray}
\label{eq:econ1}
\omega  &=&  \frac {\bfp^2_1} {2 m}   
         + \frac {\bfp^2_2} {2 m}   
         - \epsilon_{\hon}-\epsilon_{\htw} 
         + \frac {\bfp^2_r} {2 M_{A-2}}  \, , \\ 
\label{eq:qcon2}
\bqu &=& \bfp_1 + \bfp_2 + \bfp_r
\,\,\, ,
\end{eqnarray}
where, we have considered non relativistic kinematics, and we have
indicated with $\epsilon_{\hon}$ and $\epsilon_{\htw}$ the single
particle energies of the levels where the two protons are emitted
from, and with $\bfp_r$ the recoil momentum of the remaining
nucleus. The other variables have been defined in the previous
section.

\begin{table}[b]
\begin{center}
\begin{tabular}{cccc}
\hline\hline
         &   $^{12}$C     &  $^{16}$O      & $^{40}$Ca \\
\hline
 $0_1^+$ & (1p3/2)$^{-2}$ & (1p1/2)$^{-2}$ & (1d3/2)$^{-2}$ \\
 $0_2^+$ &                & (1p3/2)$^{-2}$ & (2s1/2)$^{-2}$ \\
 $1^+$   &                & (1p1/2)$^{-1}$  (1p3/2)$^{-1}$ &  
                                (1d3/2)$^{-1}$ (2s1/2)$^{-1}$\\
 $2_1^+$ & (1p3/2)$^{-2}$ & (1p1/2)$^{-1}$  (1p3/2)$^{-1}$ &
                                           (1d3/2)$^{-2}$ \\
 $2_2^+$ &                & (1p3/2)$^{-2}$ &  (1d3/2)$^{-1}$
                                              (2s1/2)$^{-1}$  \\ 
\hline\hline
\end{tabular}
\end{center}
%\vskip 0.5 cm
\caption{\small Two-hole compositions of the nuclear final states for
        the ($\gamma$,pp) reactions we have considered.  }
\label{tab:states}
\end{table}

Our calculations have been done for fixed values of $\omega$, $\bqu$
and $\epsilon_2=\bfp^2_2 / 2m$ and $\theta_2$
By selecting the final state of the rest nucleus, also
$\epsilon_{\hon}$ and $\epsilon_{\htw}$ are fixed. 
The labels
indicating the two-hole compositions of the nuclear final states of
the reactions we have considered in our calculations, are given in
Table \ref{tab:states}. We calculate the cross section as a function
of $\theta_1$, and for each value of this angle,
$|\bfp_r|$ and $|\bfp_1|$ are obtained by solving
the system formed by Eqs. (\ref{eq:econ1}) and (\ref{eq:qcon2}), this
last one squared. 
This calculation is quite involved since it should be repeated, for
every $\theta_1$, with a different single particle wave function which
depends from  $\epsilon_1=\bfp^2_1 / 2m_1$.
To relieve the numerical
load we used an approximation, called no-recoil approximation (NRA)
in \cite{ang03}.  The NRA consists in considering the mass of the
A-2 nucleus heavy enough to neglect its recoiling energy in
Eq. (\ref{eq:econ1}). In this case we have:
\begin{equation} 
|\bfp_1|=\sqrt{2m (\omega-\epsilon_2+\epsilon_{\hon}+\epsilon_{\htw})}
\end{equation}
independent from $\theta_1$. The value of  $|\bfp_r|$ is then
extracted by squaring Eq. (\ref{eq:qcon2}). 
 
The validity of the NRA has been discussed in \cite{ang03}, where we
have shown that significant differences between the results of the
exact calculations and those obtained within the NRA appear only in
the minima of the cross section angular distribution. We have obtained
analogous results for the ($\gamma$,pp) reactions.  For the purposes of
the present investigation the accuracy of NRA is sufficient, while an
eventual comparison with experimental data would require the more
accurate treatment of the kinematics.

As in \cite{ang03}, the hole single particle wavefunctions are
obtained by using a real Woods-Saxon potential whose parameters have
been fixed to reproduce charge density distributions and single
particle energies close to the Fermi level.  There is a certain degree
of arbitrariness in this choice, since there are various
parameterizations able to describe these data with the same degree of
accuracy. We have verified that this uncertainty produces on our
($\gamma$,pp) cross sections variations of few percent. This is the
same order of magnitude of the uncertainty found for the (e,e'pp) case
\cite{ang03}.

\begin{figure}[hb]
\begin{center}
\includegraphics*[width=12cm]{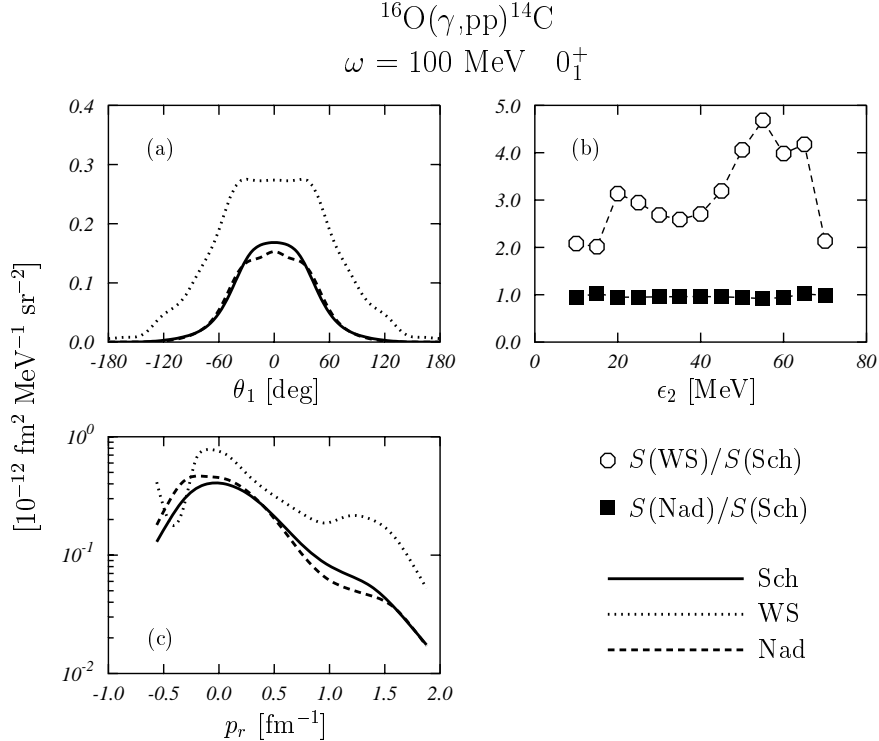}
\end{center}
\vspace*{-.5cm}
\caption{\small Two-nucleon emission cross sections on $^{16}$O
  leading to the $^{14}$C ground state.  Panel (a): Dependence on the
  angle $\theta_1$ of the first emitted proton, while the energy
  $\epsilon_2$ of the other proton is fixed at 40 MeV and
  $\theta_2=180^o$. The full line shows the result obtained with the
  optical potential of Schwandt et al.  \cite{sch82}, the dashed line
  with that of Nadasen et al. \cite{nad81}, while the dotted line has
  been obtained by using for the particle states the same real
  Woods-Saxon well used to generate the hole single particle wave
  functions.  Panel (b): Ratios between $S$ variables, Eq.
  (\protect\ref{eq:ratx}) calculated with real Woods-Saxon well (WS),
  Nadasen potential (Nad) and Scwhandt potential (Sch). The lines have
  been drawn to guide the eyes.  Panel (c): Cross sections in
  superparallel back-to-back kinematics. The variable $|\bfp_r|$ has
  been modified by changing the energies of the emitted nucleons, and
  it has been defined positive when it points in the direction of the
  momentum transfer $\bqu$. The meaning of the lines is the same as in
  the panel (a).}
\label{fig:opt}
\end{figure}

The continuum single particle wave functions, have been described by
using the complex optical potential of \cite{sch82}, which was fixed
to reproduce elastic nucleon-nucleus cross section data. In our case
this complex potential takes into account the re-scattering of the
emitted nucleon with the rest nucleus. The interaction between the two
emitted nucleons is neglected. Recent calculations \cite{sch04}
indicate that, for photons and for the kinematics we have chosen, this
effect is negligible. The uncertainty on the choice of the continuum
wave functions produce large effects on the cross section, and this
fact deserves a more detailed discussion than for the bound wave
functions.  In Fig. \ref{fig:opt} we show the results of the
$^{16}$O($\gamma$,pp)$^{14}$C cross sections calculated with
different sets of continuum wave functions, for the final state
$0^+_1$ (see Table \ref{tab:states}). In the panel (a) of this figure,
we show the cross sections as a function of the $\theta_1$ angle. In
these calculations the energy and the angle of the other emitted
proton have been fixed as $\epsilon_2$ = 40 MeV and
$\theta_2=180^o$. The full line shows the result obtained by using the
wavefunctions obtained with the optical potential of Schwandt et
al. \cite{sch82}.  This is the potential we have used all along the
remaining of the paper. A different parameterization of an optical
potential also aiming to reproduce elastic scattering proton-nucleus
data, is that of Nadasen et al. \cite{nad81}. The dashed line of panel
(a) shows the result obtained with this optical potential. The dotted
line shows the result obtained by using, for the particle states, the
same real Woods-Saxon potential adopted for the hole states. The
parameters of this potential can be found in \cite{ari96}.

As it is well know, the imaginary part of the optical potential
reduces the cross section. The use of these potentials is necessary to
have a good description of single nucleon emission processes
\cite{bof96}. In this approach the hole and particle wave functions
are not any more orthogonal. The effects of this non-orthogonality
have been investigated in \cite{bof82} and have been found to be small
in the kinematic region we consider.

The number of the kinematic variables involved in the cross section
calculations is quite large, therefore we checked the range of
validity of our results by repeating the calculations under different
kinematic conditions.
Quite often the details of the angular distribution of the cross
section are not relevant for our discussion We found useful to study
the integrated cross section defined as:
\beq
S\, = \, \int {\rm d}\theta_1 \sin \theta_1
 \frac{{\rm d}^5 \sigma(\theta_1)}
{ {\rm d}\Omega_1
{\rm d}\epsilon_2 {\rm d}\Omega_2 } \, .
\label{eq:ratx}
\eeq
which summarizes the global differences between the various
calculations.

The integrated cross sections $S$ have been calculated with the sets
of single particle wave functions discussed above for $\theta_2=180^o$
and various values of the energy $\epsilon_2$.  The results of these
calculations are shown in the panel (b) of Fig. \ref{fig:opt} as a
ratios of the $S$ variables calculated with Nadasen and real
Woods-Saxon potential divided by those obtained with the Schwandt
potential.  It is evident that the behaviour shown in panel (a) is a
general feature of the calculation. FSI reduce the cross
sections, and the two optical potentials produce very similar results.

We have also investigated the effects of special kinematics on the
FSI. Specifically we have chosen the super parallel back-to-back
kinematics. In this case one proton is emitted in the direction of
the momentum transfer and the other one in the opposite direction. We
set $\theta_1=0^o$ and $\theta_2=180^o$. The kinematics is changed by
modifying the energies of the emitted protons. An easy way to
represent the cross sections in this case is to use $|\bfp_r|$ defined
in Eq. (\ref{eq:qcon2}). For fixed emission angles, the same value of
this quantity can be obtained for two sets of energy values of the two
emitted protons. In the superparallel back-to-back kinematics we can
distinguish when $\bfp_r$ points in the direction of the momentum
transfer $\bqu$ or in opposite direction. We decided to assign
positive values to $|\bfp_r|$ in the first case, and negative in the
second one. For other kinematic set ups the use of $|\bfp_r|$ is
more ambiguous.

The panel (c) of Fig. \ref{fig:opt} shows the cross sections
calculated for the kinematics described above with the potentials we
are discussing. As expected the real Woods-Saxon potential produce
larger cross sections, while the two optical potentials produce very
similar cross section. We have calculated the values of the cross
sections integrated in $\bfp_r$ and taken the ratios. We found a value
of 2.06 for WS over Schwandt and 1.09 for Nadasen over Schwandt. The
first value should be compared with the open circles of panel (b), and it
is of the same order of the smallest values shown in the figure. The
other value should instead be compared with the squares of panel
(b). It is slightly larger than those shown. These results do not
present any indication that FSI are reduced in superparallel
back-to-back kinematics.

This investigation has been conducted also for the other $^{16}$O
states listed in the Table \ref{tab:states}, and we found similar
results. We can summarize our results by saying that, if one uses
potentials that reproduce ground state properties and nucleon-nucleus
elastic scattering data, the uncertainties related to the choice of
single particle wave functions is about the 10\%.

\subsection{The $\Delta$ currents}
The two-nucleon emission induced by two-body currents, for inclusive
electron scattering processes, has been investigated in \cite{ama94}.
The results presented in that article indicate that the emission of
two-like nucleons is a small contribution to the total, inclusive,
cross section. In our case, however, it is important to know the
relevance of the two-proton emission induced by $\Delta$ currents with
respect to that induced by the SRC.  There are two questions we would
like to address in this part of the paper.  The first one is about our
capacity of describing the $\Delta$ currents, and the second one is
about the possibility of finding particular kinematics able to minimize
their effects.

\begin{figure}
\begin{center}
\includegraphics*[width=11cm]{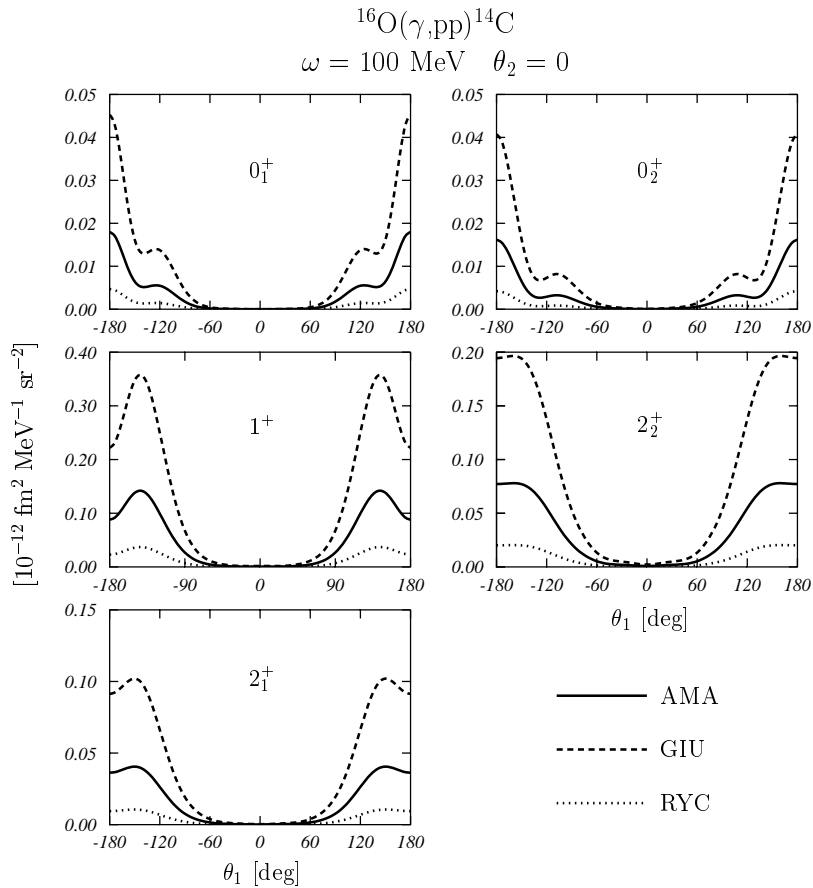}
\end{center}
\vspace*{-.5cm}
\caption{\small Cross sections angular distributions, as a function of
  $\theta_1$, calculated with different parameterizations of the
  $\Delta$ coupling constants (see Table \protect\ref{tab:delta}). 
  The calculations have been done considering the $\Delta$ currents
  only and with  $\epsilon_2$=40 MeV. The labels of the various panels
  indicate the final states.}
\label{fig:delta}
\end{figure}

In Fig. \ref{fig:delta} we show the $^{16}$O($\gamma$,pp)$^{14}$C
cross sections as a function of the $\theta_1$ angle, for the final
states indicated in Table \ref{tab:states}. The cross sections have
been calculated for $\omega$=100~MeV, $\epsilon_2$=40~MeV, and
$\theta_2= 0^o$. The calculations have been done considering only the
$\Delta$ currents, i.e. there is not interference with the
contribution of the SRC.  The three curves of each panel have been
obtained with different values of the $f_{\gamma N \Delta}$ and
$f_{\pi N \Delta}$ coupling constants used in the evaluation of the
diagrams of Fig. \ref{fig:feydelta}. A discussion of the validity of
the various parameterizations is out of the scope of our work, we
simply want to investigate the uncertainty produced on our results.
The values we have used are taken from \cite{ama94} (AMA),
\cite{giu97} (GIU), \cite{ryc97} (RYC), and they are given in Table
\ref{tab:delta}. While the shapes of the angular distributions are not
affected, the sizes of the cross sections are strongly modified by the
various choices of the parameters.

\begin{table}[b]
\begin{center}
\begin{tabular}{rrrr}
\hline\hline
                            & AMA & GIU & RYC  \\
\hline
 $f_{\gamma N \Delta}$      & 0.299 & 0.373 & 0.120  \\
 $f_{\pi N \Delta}$         & 1.69  & 2.15  & 2.15  \\
\hline\hline
\end{tabular}
\end{center}
%\vskip 0.5 cm
\caption{\small Values of the parameters used in the evaluation of the
  $\Delta$ currents contribution.  The AMA, GIU and RYC values are
  from Refs. \protect\cite{ama94}, \protect\cite{giu97} and
  \protect\cite{ryc97}, respectively.  }
\label{tab:delta}
\end{table}
\begin{figure}
\begin{center}
\includegraphics*[width=11cm]{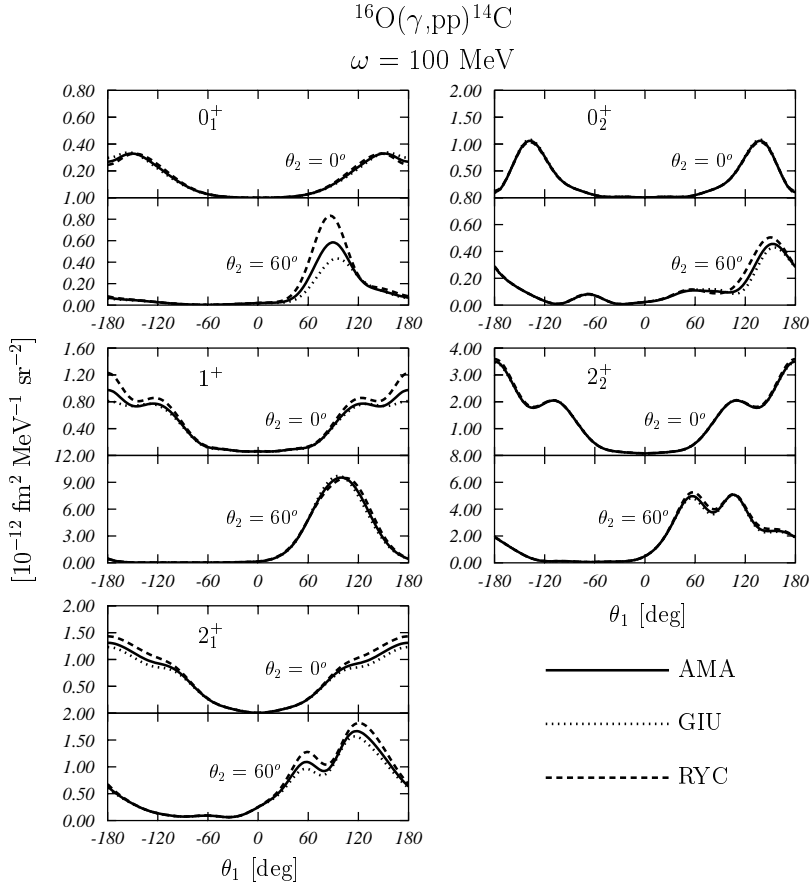}
\end{center}
\vspace*{-.5cm}
\caption{\small Cross sections angular distributions, as a function of
  $\theta_1$, calculated with different parameterizations of the
  $\Delta$ coupling constants (see Table \protect\ref{tab:delta}). 
  All the calculations have
  been done with $\epsilon_2$=40 MeV. The labels of the various panels
  indicate the final states and the values of $\theta_2$.}
\label{fig:giu}
\end{figure}
\begin{figure}
\begin{center}
\includegraphics*[width=11cm]{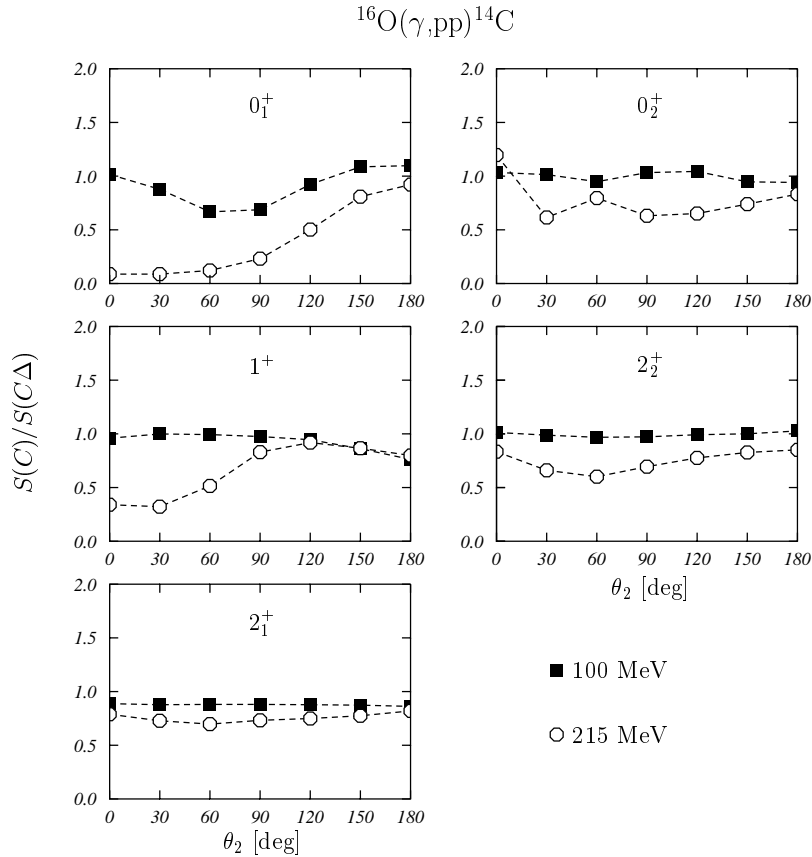}
\end{center}
\vspace*{-.5cm}
\caption{\small Ratio $S(C)/S(C\Delta)$, see Eq.
       (\protect\ref{eq:ratx}), calculated for various values of
       $\theta_2$ and for photon energies of 100 and 215 MeV. The
       calculations have been done with $\epsilon_2$=40 MeV.
       The dashed lines have been drawn to guide the eyes.  }
\label{fig:int215}
\end{figure}
\begin{figure}
\begin{center}
\includegraphics*[width=11cm]{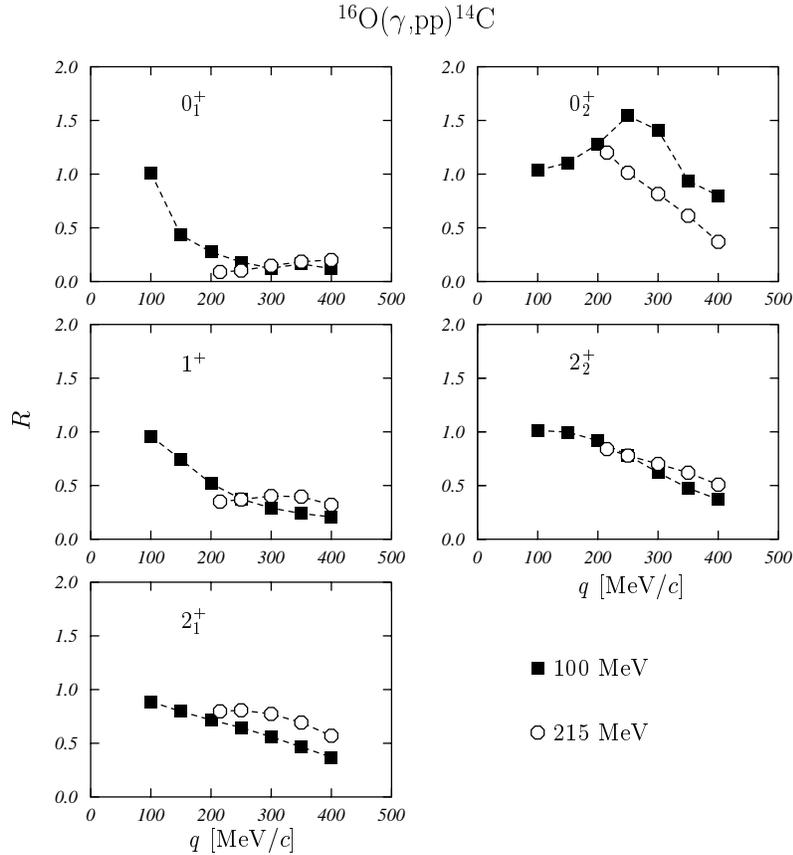}
\end{center}
\vspace*{-.5cm}
\caption{\small Ratio between the integrated $w_t(|\bqu|)$ responses
       calculated without and with $\Delta$ currents, see Eq.
       (\protect\ref{eq:R}). The ratios have
       been calculated for $\theta_2=0^o$.}
\label{fig:intq}
\end{figure}
\begin{figure}
\begin{center}
\includegraphics*[width=11cm]{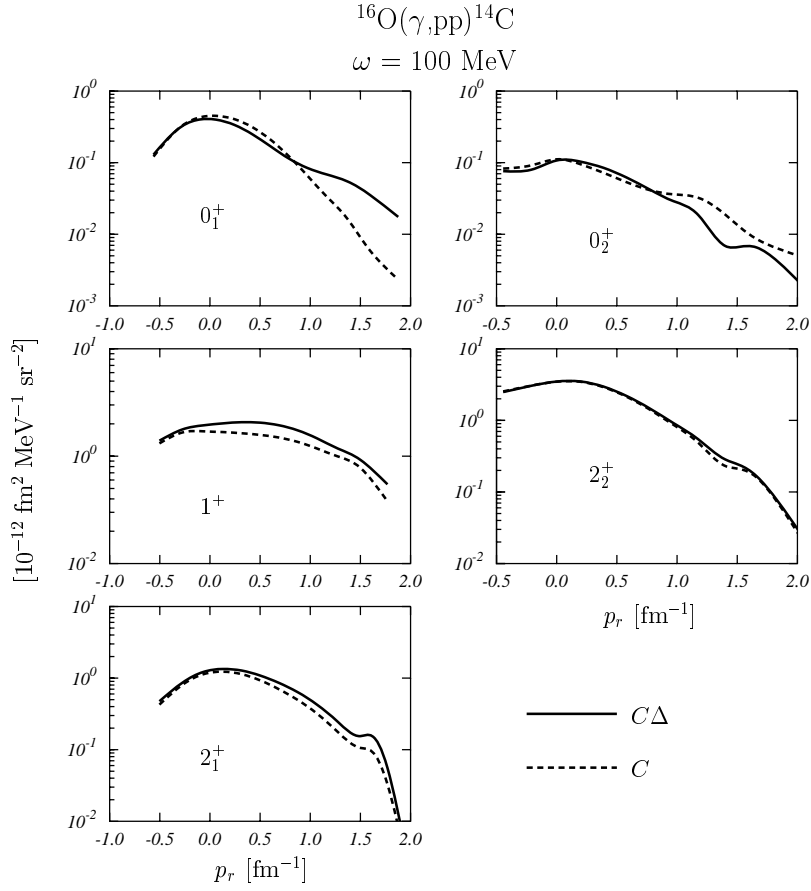}
\end{center}
\vspace*{-.5cm}
\caption{\small Cross sections in superparallel back to back
  kinematics with $\theta_2=180^o$.
  The variable $|\bfp_r|$ has been defined, and changed,
  as in the panel 
  (c) of Fig. \protect\ref{fig:opt}. The full lines have been obtained
  by considering both SRC and $\Delta$ terms, while the dashed lines
  show the results obtained without the  $\Delta$ terms.}
\label{fig:spdelta}
\end{figure}
\begin{figure}
\begin{center}
\includegraphics*[width=11cm]{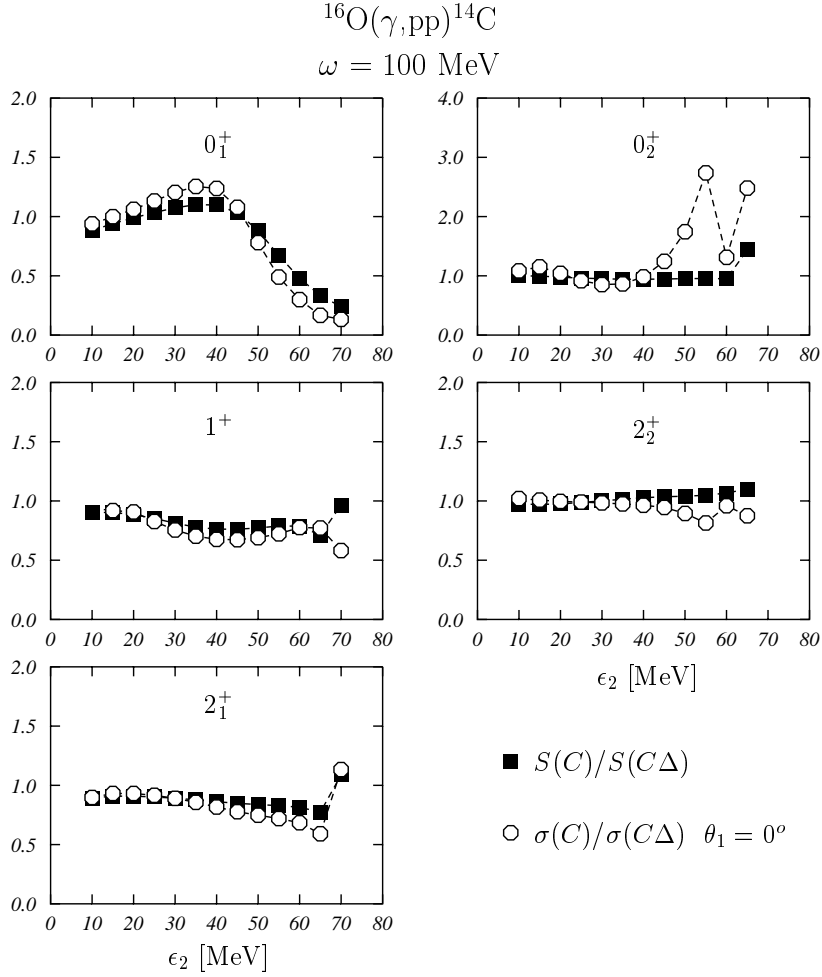}
\end{center}
\vspace*{-.5cm}
\caption{\small The squares show the ratios between the S factors of
  Eq. (\protect\ref{eq:ratx}) calculated without and with $\Delta$
  currents respectively.  The circles show the ratios of the cross
  sections of superparallel kinematics shown in Fig.
  \protect\ref{fig:spdelta} calculated with and without delta. The
  ratios are shown as a function of the energy $\epsilon_2$, and they
  have been calculated with $\theta_2=180^{o}$. }
\label{fig:intsp}
\end{figure}
\begin{figure}
\begin{center}
\includegraphics*[width=12cm]{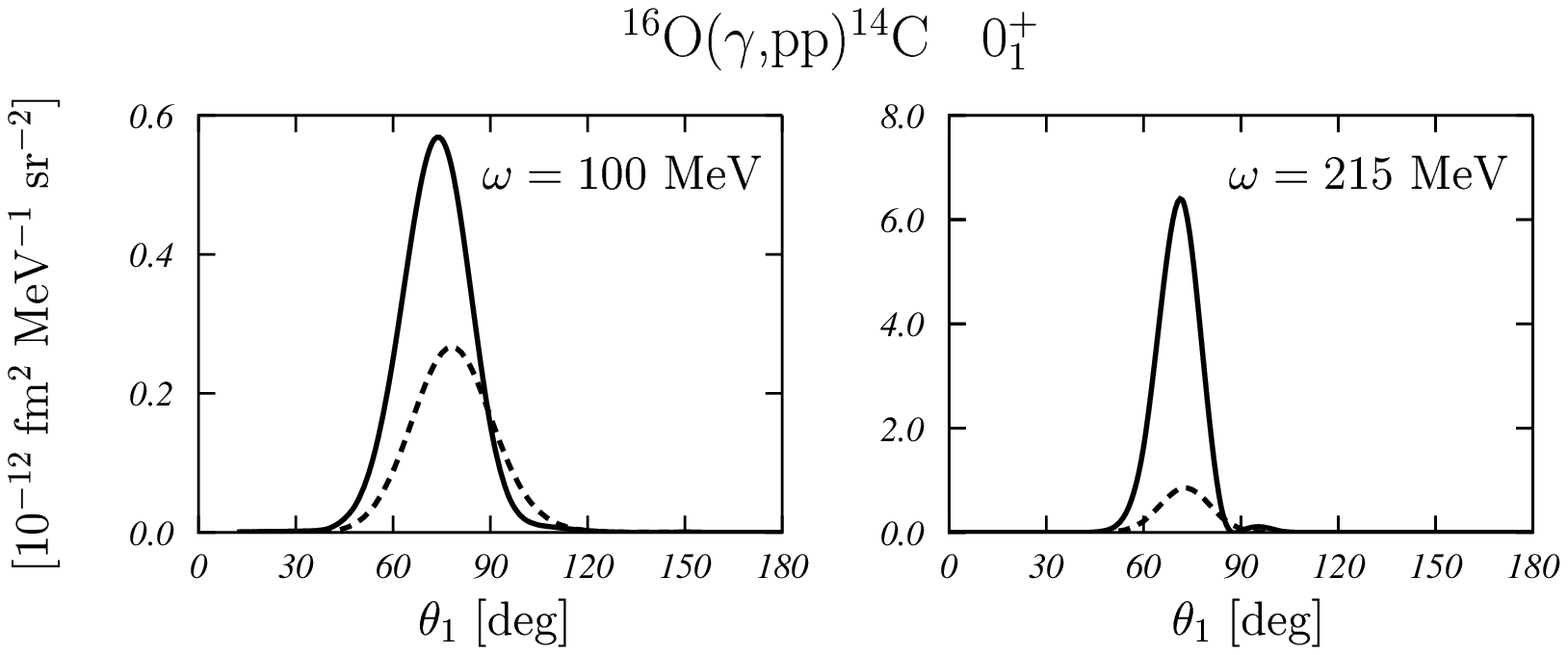}
\end{center}
\vspace*{-.5cm}
\caption{\small Cross section angular distributions for the $0^+_1$
  final state in $^{14}$C in symmetric kinematics (see
  sect. \ref{sec:correl}).  
  The full lines
  show the results obtained by considering both SRC and
  $\Delta$-currents. The results shown by the dashed lines have been
  obtained with SRC only.
  }
\label{fig:symm}
\end{figure}

The uncertainty on the global result depends from the interference
between $\Delta$ currents and SRC. In Fig. \ref{fig:giu} we show the
cross section angular distributions for the same kinematics of Fig.
\ref{fig:delta}, and also for $\theta_2= 60^o$, when the SRC are
considered (the correlation function is the gaussian correlation we
shall present in section \ref{sec:correl}). The results of Fig.
\ref{fig:giu} show that the differences between the various curves are
rather small, apart from the case of $0_1^+$ state at 60$^o$.
By integrating in $\theta_1$ the cross sections, as indicated in Eq.
(\ref{eq:ratx}), we obtain a maximum difference of 10\%, between the
various calculations, with the exclusion of the case mentioned above 
where the differences can reach the 45\%.

As indicated in \cite{ama94} the effects of the $\Delta$ currents
increase when the excitation energy becomes closer to the peak of the
of the $\Delta$ resonance. This is evident from Fig.
\ref{fig:int215} where we show the ratios between the $\theta_1$
integrated cross sections $S$ defined in Eq. (\ref{eq:ratx}),
calculated by using SRC only, $S(C)$, and by including also the
$\Delta$ currents $S(C\Delta)$.  The process investigated is
$^{16}$O($\gamma$,pp)$^{14}$C, calculated for various values of
$\theta_2$ by using $\epsilon_2$=40~MeV, for different final states of
$^{14}$C and for two different photon energies, 100 and 215~MeV.
These two energy values have been chosen to be far from the giant
resonance region, to minimize collective phenomena (see the discussion
in \cite{ang02}).

The smaller is the ratio the larger is the effect of the $\Delta$
currents. The general behaviour of the results shown in the figure
confirms what was expected, i.e. the effects of the $\Delta$ currents
are smaller at 100 MeV than at 215 MeV. More quantitative statements
are difficult to make because of the strong dependence of the results
from both the final state and $\theta_2$. 
The behaviour of  the 0$^+_1$ and
1$^+$ results at $\omega$ = 215 MeV are quite anomalous with
respect to those of 
the other results at the same energy. It is also worth to
notice, in the case of $\omega$=100 MeV, the peculiar results for the
0$^+_1$ final state at 60$^o$ and 90$^o$.
 
Apart from the cases mentioned above, all the ratios calculated at
$\omega$ = 100 MeV are very close to the unity, showing a small effect
of the $\Delta$ currents. The study of (e,e'pp) reactions done in
\cite{ang03} shows larger effects of the $\Delta$ currents, especially
if one consider the case of the 1$^+$ state. A possible source of this
difference can be related to the different values of the momentum
transfer in photo and electron reactions. We have calculated the
transverse response of Eq. (\ref{eq:wt}) for various values of
$|\bqu|$. In Fig. \ref{fig:intq} we show the ratio
\beq
R\, = \, \displaystyle 
\frac{\displaystyle \int {\rm d}\theta_1 \sin \theta_1 w_t^C(|\bqu|)}
{\displaystyle \int {\rm d}\theta_1 \sin \theta_1
w_t^{C\Delta}(|\bqu|)} 
\label{eq:R}
\eeq
for the excitation energies and states considered before. The responses
$w_t^C$ have been calculated with SRC only, and $w_t^{C\Delta}$ by
including also the $\Delta$ currents.  The calculations have been done
for $\theta_2=0^o$ and $\epsilon_2$=40 MeV.

The figure shows that the contribution of the $\Delta$ currents
increases with increasing value of $|\bqu|$. Also this general trend
has an exception: the $0^+_2$ at $\omega$ = 100 MeV. It is however
relevant the fact that, for all the considered cases, the value of the
ratio at $\omega=100$ MeV is very close to unity at the photon point.
This indicates that the two-body $\Delta$ currents do not contribute
significantly for photons.

In searching for kinematics set ups where the $\Delta$ currents
contributions are rather small, we have investigated with some detail
the case of the superparallel back-to-back case, which is one of
the preferred situations from the experimental point of view.  In
Fig. \ref{fig:spdelta} we show the cross sections for the
$^{16}$O($\gamma$,pp)$^{14}$C reactions. The calculations have been
done for $\omega$ = 100 MeV, $\theta_2=180^{o}$ and $\theta_1=0^{o}$.
The value of $|\bfp_r|$ has been defined as in Fig. \ref{fig:opt},
and it has been changed by modifying $\epsilon_2$. The dashed
lines have been obtained with SRC only, and the full lines contain
also the $\Delta$ currents.  The results of the two calculations are
rather similar, showing again the scarce relevance of the $\Delta$
currents at $\omega$=100 MeV.  The only remarkable difference between
calculations with and without $\Delta$ currents appears for the
$0^+_1$ final state in the region of the high positive values of
$|\bfp_r|$.

These conclusions are not a special feature related to the chosen
kinematics but they represent a general trend.  In Fig.
\ref{fig:intsp} the squares show the ratios between the $S$ factors
(Eq. \protect\ref{eq:ratx}) calculated without and with $\Delta$
currents as a function of $\epsilon_2$ (black squares). In the same
figure, the ratios between the superparallel cross sections of Fig.
\protect\ref{fig:spdelta}, calculated without and with $\Delta$
currents are shown by the open circles. In general the two ratios are
very similar, and this indicates that the $\Delta$ currents are not
suppressed in superparallel kinematics.  Remarkable the behaviour of
the $0^+_2$ for large $\epsilon_2$ values. This is related to some
peculiarity of the superparallel kinematics as we shall discuss in the
next section.

The results we have just discussed do not imply that the $\Delta$
currents are always negligible.  It possible to find specific
kinematics where they become more important than the SRC.  As example,
we show in Fig. \ref{fig:symm} the results obtained for the $0^+_1$
state for the $^{16}$O($\gamma$,pp)$^{14}$C reaction in symmetric
kinematics.  This kinematics consists in setting $\theta_1=\theta_2$,
and $\epsilon_1=\epsilon_2=0.5(\omega
-\epsilon_{\hon}-\epsilon_{\htw})$.  The figure shows that, in this
case, the $\Delta$ currents play an important role at both photon
energies considered.

\subsection{The correlations}
\label{sec:correl}
The main goal in the study of the two nucleon emission processes
regards the possibility of obtaining information about SRC. We
conducted this investigation by comparing the results obtained with
different SRC.

All the results presented up to now have been obtained by using a
gaussian correlation of the type
\beq
f(r)\, = \, 1 \,- \,a \, \exp \left( -b r^2 \right) \, ,
\eeq
with $a=0.7$ and $b=2.2~{\rm fm}^{-2}$. These values have been fixed
in \cite{ari96} by minimizing the energy functional for nuclear
hamiltonian containing the Afnan and Tang nucleon-nucleon interaction
\cite{afn68}. This calculation has been done in the framework of the
Correlated Basis Function theory by using Fermi Hypernetted Chain
resummation techniques.

\begin{figure}
\begin{center}
\includegraphics*[width=7cm]{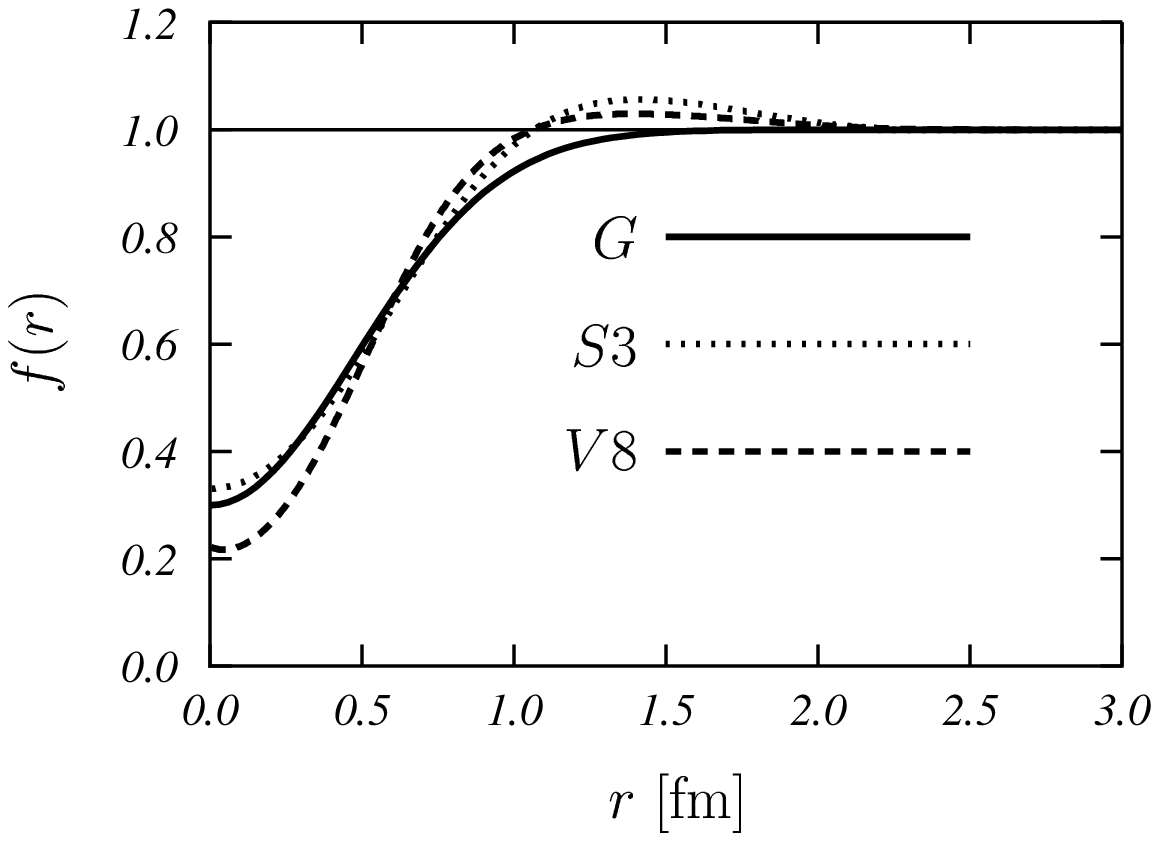}
\end{center}
\vspace*{-.5cm}
\caption{\small Correlation functions used in our calculations as a
  function the relative distance between the nucleons. The line labeled $G$
  and $S3$ indicate the gaussian and Euler correlations of
  \cite{ari96}. With $V8$ we show the scalar term of the
  state-dependent correlation of \cite{fab00}. }
\label{fig:corr}
\end{figure}

Within the same theoretical approach, 
the other two correlations have been fixed by using a method called
Euler
procedure \cite{co92}. With this procedure the SRC depend only from
a single parameter, the healing distance, whose value 
is also fixed by a
minimization the energy functional.  The correlation function labeled
$S3$ in the figure is obtained in this way \cite{ari96} when the Afnan
and Tang nucleon-nucleon interaction is used in Fermi Hypernetted
Chain calculations of the doubly magic nuclei $^{12}$C, $^{16}$O,
$^{40}$Ca, $^{48}$Ca and $^{208}$Pb.
The other correlation function we have adopted in our calculation,
labeled $V8$ in the figure, is the scalar term of the state dependent
correlation used in \cite{fab00} for the Fermi Hypernetted Chain
calculations of the $^{16}$O and $^{40}$Ca nuclei done with the V8'
parameterization of the realistic Argonne V18 nucleon-nucleon
potential \cite{wir95} plus the Urbana IX three-body interaction
\cite{pud97}.
As it is shown in Fig. \ref{fig:corr} the three SRC differ for few
details. The $V8$ and $S3$ correlations become larger than the
asymptotic value of 1 in the region between 1 and 2 fm. The $V8$
correlation has a lower minimum than the other two.

\begin{figure}
\begin{center}
\includegraphics*[width=12cm]{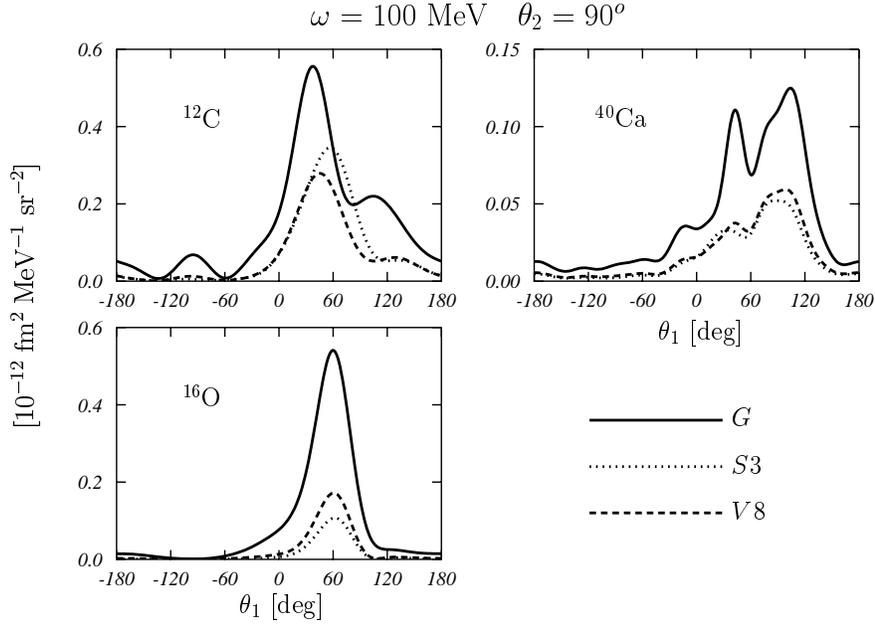}
\end{center}
\vspace*{-.5cm}
\caption{\small Angular distributions of the ($\gamma$,pp) 
  cross sections for the three different nuclei considered. The
  kinematics variables have been fixed as 
  $\omega=100$ MeV, $\theta_2=90^o$ and $\epsilon_2$=40 MeV. The
  final states are the ground states of the A-2 nuclei, which
  correspond to the $0^+_1$ states of Table \protect\ref{tab:states}.  
  The full lines have been obtained with
  the gaussian correlation, the dotted ones with the $S3$ correlation
  and the dashed lines with the $V8$ correlation.
  }
\label{fig:angc}
\end{figure}

We have studied the ($\gamma$,pp) reaction for $^{12}$C, $^{16}$O and
$^{40}$Ca target nuclei. As example of the obtained results, we show
in Fig. \ref{fig:angc} some cross section angular distributions.  The
results of the figure have been obtained for final states leading to
the ground state of the A-2 nuclei, with $\theta_2$=90$^o$ and
$\epsilon_2$=40 MeV.  The shapes of the angular distributions are not
strongly modified by the various correlations. The main difference
between the various results is in the size of the cross section.

\begin{figure}
\begin{center}
\includegraphics*[width=11cm]{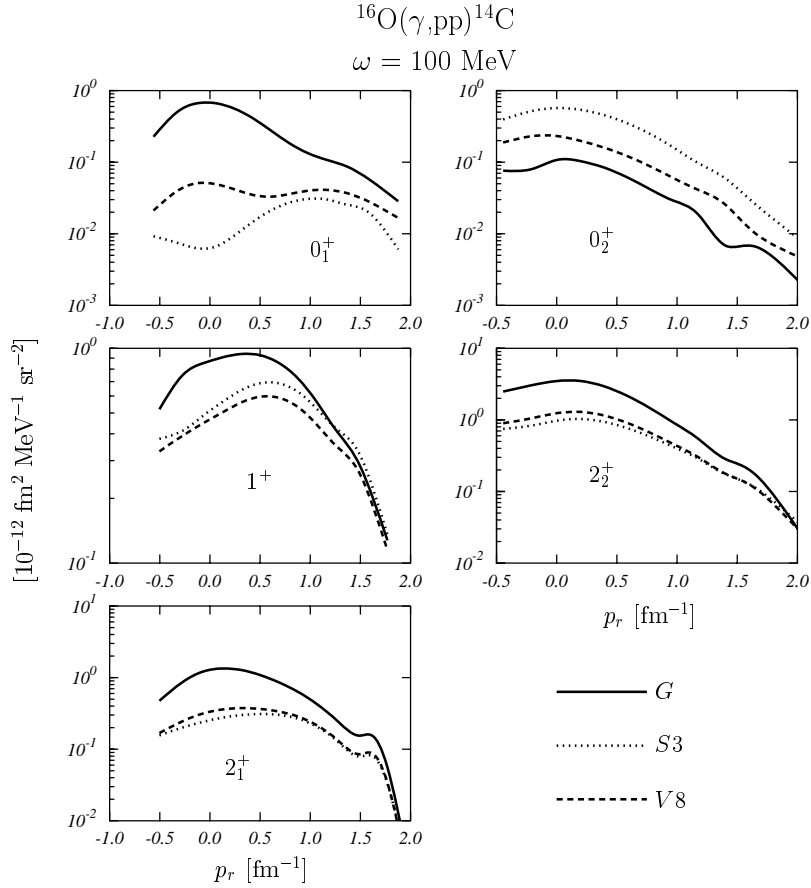}
\end{center}
\vspace*{-.5cm}
\caption{\small $^{16}$O($\gamma$,pp)$^{14}$C  cross sections
  calculated  for various final states in superparallel back to back
  kinematics ($\theta_1=0^o$,$\theta_2=180^o$). The variable $|\bfp_r|$
  has been defined, and changed, as in Fig. \ref{fig:opt}.
  }
\label{fig:sp}
\end{figure}
\begin{figure}
\begin{center}
\includegraphics*[width=11cm]{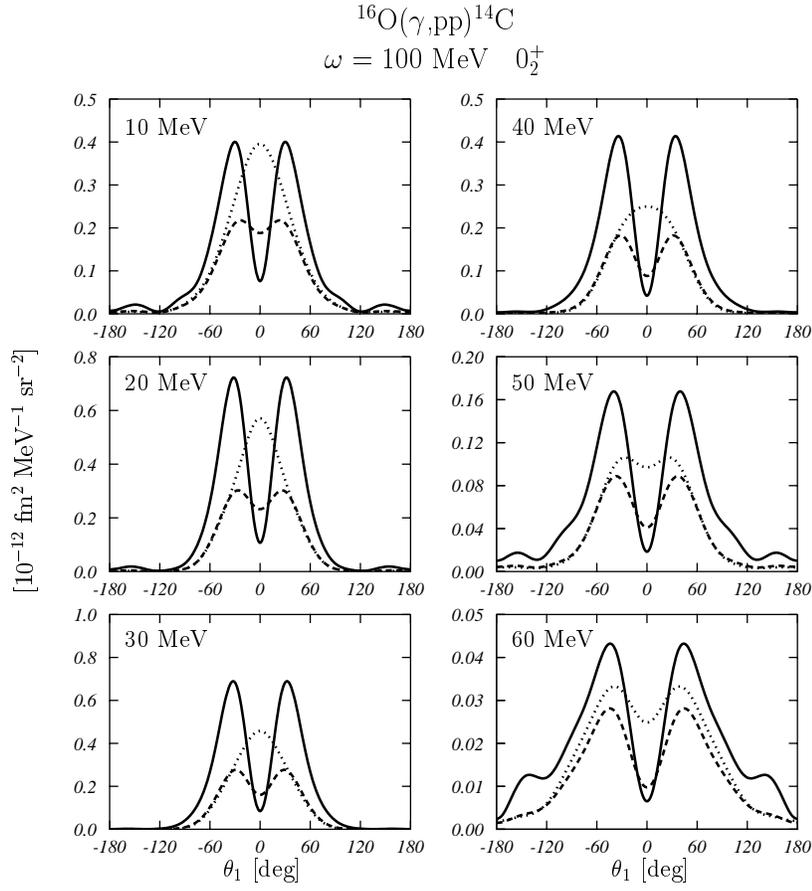}
\end{center}
\vspace*{-.5cm}
\caption{\small Angular distributions of the
$^{16}$O($\gamma$,pp)$^{14}$C cross sections for the $0^+_2$ final
state. The energy values in the various panels correspond to
$\epsilon_2$. The meaning of the lines is analogous to that of
Fig. \ref{fig:sp}: full, dotted and dashed lines correspond to the $G$,
$S3$ and $V8$ correlations, respectively.}
\label{fig:220p}
\end{figure}
\begin{figure}
\begin{center}
\includegraphics*[width=11cm]{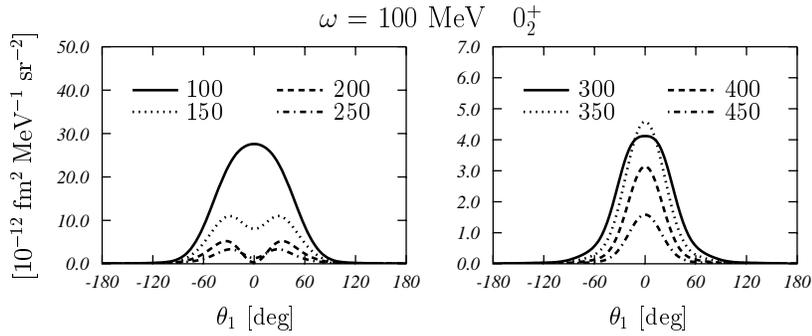}
\end{center}
\vspace*{-.5cm}
\caption{\small Response functions $w_t(\bqu)/|\bfp_1|^2|\bfp_2|^2$
  for the emission of two protons from $^{16}$O calculated with the $S3$
  correlation function for $\theta_2=180^o$ and $\epsilon_2$ = 40 MeV.
  The labels explaining the meaning of the lines indicate the values
  of $|\bqu|$ in MeV/c.  }
\label{fig:220pq}
\end{figure}
\begin{table}[b]
\begin{center}
\begin{tabular}{ccc cc cc}
\hline\hline
         &\multicolumn{2}{c } {$^{12}$C}     &  
         \multicolumn{2} {c } {$^{16}$O}      & 
         \multicolumn{2} {c} {$^{40}$Ca} \\
\hline
     & $S3$ & $V8$ & $S3$ & $V8$ & $S3$ & $V8$ \\
\hline
 $0_1^+$  & 0.73 $\pm$ 0.11 & 0.52 $\pm$ 0.05  
          & 0.10 $\pm$ 0.11 & 0.20 $\pm$ 0.05 
          & 0.15 $\pm$ 0.03 & 0.22 $\pm$ 0.04  \\
 $0_2^+$ & &   
         & 0.83 $\pm$ 0.16 & 0.60 $\pm$ 0.08 
         & 0.37 $\pm$ 0.07 & 0.44 $\pm$ 0.07  \\
 $1^+$   & &   
         & 0.97 $\pm$ 0.08 & 0.76 $\pm$ 0.07
         & 0.37 $\pm$ 0.04 & 0.36 $\pm$ 0.06  \\
 $2_1^+$ & 0.56 $\pm$ 0.13 & 0.52 $\pm$ 0.08  
         & 0.25 $\pm$ 0.05 & 0.30 $\pm$ 0.04 
         & 0.22 $\pm$ 0.05 & 0.24 $\pm$ 0.03  \\
 $2_2^+$ & &   
         & 0.54 $\pm$ 0.13 & 0.52  $\pm$ 0.08 
         & 0.54 $\pm$ 0.07 & 0.48 $\pm$ 0.05  \\
\hline\hline
\end{tabular}
\end{center}
%\vskip 0.5 cm
\caption{\small Ratios between the $S$ factors calculated with the
  $S3$ and $V8$ correlations and those calculated with the Gaussian
  correlation, averaged on the $\theta_2$ variable (see sect.
  \ref{sec:correl}). }
\label{tab:corrat}
\end{table}

To have concise information from our calculations we have
evaluated the $S$ factors of Eq. (\ref{eq:ratx}), for all the
states of Table \ref{tab:states}, for $\epsilon_2$=40 MeV and for
$\theta_2=0^o$, 30$^o$, 60$^o$, 90$^o$, 120$^o$, 150$^o$ and
180$^o$. We show in Table \ref{tab:corrat} the ratios between the $S$
factors calculated with $S3$ and $V8$ and those calculated with the
gaussian correlation averaged on the $\theta_2$ variable. In the table
the uncertainty on the average is calculated as a standard deviation.

All the values given in the table are smaller than one, and this
indicates that the gaussian correlation produces the largest cross
sections.  In general, the $V8$ cross sections are larger than those
obtained with the $S3$ correlation.  Analogous results have been
obtained in \cite{ang03}.  In that reference, it has been shown that
the behaviour of the $S3$ and $V8$ correlation functions in the region of
1-2 fm where they assume values larger than one, has opposite effect
with respect to that of the region at smaller $r$ values. Gaussian and
$S3$ correlation functions have the same minimum, but the gaussian
correlation does not have the overshooting of the asymptotic values.
For this reason the cross sections obtained with the gaussian
correlation are larger than the other ones.  The $V8$ correlation
function has a smaller minimum, and also its overshooting values in
the 1-2 fm region are smaller than those of the $S3$ correlation
function.

These general features of our results, obtained by making angular
integrations and averages, are not respected in every kinematic
situation.  The cross sections strongly depend upon all the kinematic
variables, from the nuclear final states up to the angles and the
energies of the two emitted protons. As an example of this sensitivity
we show in Fig. \ref{fig:sp} the $^{16}$O($\gamma$,pp)$^{14}$C
cross sections calculated for various final states in superparallel
back-to-back kinematics. As before, the variable $\bfp_r$ has been
changed by changing the energy $\epsilon_2$, and consequently
$\epsilon_1$.

Also in this case we observe that the general behaviour emerging from
the study of Table \ref{tab:corrat} is respected. The cross sections
calculated with the gaussian correlation are the largest ones.  The
$0^+_2$ results show an opposite behaviour. The largest cross sections
are those produced by the $S3$ correlation, and the smallest ones by
the gaussian correlation. To understand better the source of the of
this difference we show in Fig. \ref{fig:220p} the angular
distributions of the $^{16}$O($\gamma$,pp)$^{14}$C cross sections
for the $0^+_2$ final state. The cross section values used for the
superparallel back-to-back results are those at $\theta_1$=0$^o$ where
the $S3$ cross sections are larger than the other ones.  On the other
hand the figure clearly shows that the largest values of the angular
integrated cross sections are those obtained with the gaussian
correlation.

The shapes of the angular distributions shown in Fig. \ref{fig:220p}
resemble those of the (e,e'p) experiments when the emitted particle is
coming from a bound $p$ wave, with the typical minimum at the origin
\cite{bof96}. The ($\gamma$,pp) cross section is related to the wave
function describing the motion of the correlated pair in target
nucleus ground state \cite{giu98}.  This is the combined wave function
of the two-particles with respect to the nuclear center.  Clearly the
result depends also from the correlation used, as the different lines
of Fig. \ref{fig:220p} show. In any case the typical $p$ wave
behaviour is present in all the cases. Also the $S3$ results start to
show the characteristic minimum at the origin when the kinematics
allows to probe small values of the relative momentum of the
correlated hole pair. This behavior of the $S3$ response function is
also shown in Fig. \ref{fig:220pq} where the angular distributions of
the $w_t(|\bqu|)$ responses are shown as a function of $\theta_1$ for
different values of $|\bqu|$. This manner of changing the kinematics
cannot be achieved in photo emission, but in electron scattering. The
figure shows that for certain values of $|\bqu|$ the angular
distributions have a minimum at the origin.

\section{Summary and conclusions}
\label{sec:con}
In this paper we have applied to the ($\gamma$,pp) reaction the model
developed in \cite{co98,mok00,co01,mok01,ang02}.  We first analyzed
the sensitivity of our results to the uncertainties related to the
choices of the mean field potentials describing the single particle
wave functions. Hole and particle wave functions are generated by two
different potentials whose parameters are fixed by using different
criteria.  The hole wave functions should describe some property of
the target nucleus ground state, for example charge distributions and
single particle energies. The particle wave functions are generated by
optical potentials that reproduce the elastic nucleon-nucleus cross
sections.  The use of various potentials, all of them satisfying these
criteria, produces variations of few percent on the cross section.

We have conducted a study on the relevance of the $\Delta$ currents on
the cross section, since they are the only mechanism, competing with
SRC, able to emit two protons.  The uncertainty on the values of the
coupling constants produces large effects on the $\Delta$ currents
(see Fig. \ref{fig:delta}). Fortunately the contribution to the
cross section of the $\Delta$ currents is, in general, smaller than
that of the SRC. Therefore this uncertainty affects the cross section
values at the 10\% level.  Because of the large number of kinematic
variables into play this result should be  more carefully stated.
As expected, the $\Delta$ currents are small in the region far from
the peak of the $\Delta$ resonance, and become relatively larger while
approaching it (see Fig. \ref{fig:int215}). On the other hand, we
have shown in Fig. \ref{fig:intq}, that at fixed nuclear excitation
energy, the relative effect of the $\Delta$ currents increases with
increasing momentum transfer. For this reason the $\Delta$ currents
affect more the (e,e'pp) processes than the ($\gamma$,pp) reactions.

The choice of the kinematics could modify these results. We analyzed
superparallel back-to-back kinematics and found that they reproduce
the general trend obtained by integrating on the angular distribution
of one of the emitted protons (see Figs. \ref{fig:spdelta} and 
\ref{fig:intsp}). On the other hand, in a different kinematics, the
symmetric one, the  $\Delta$ currents become more important than the
SRC (see Fig. \ref{fig:symm}). 

The sensitivity of the cross section to the details of the correlation
has been studied by using three different correlation functions, shown
in Fig. \ref{fig:corr}. The calculations have been done for the 
$^{12}$C, $^{16}$O and $^{40}$Ca nuclei. The various correlations do
not sensitively modify the shapes of the angular distributions, but
they modify the the size of the cross section. We observe that the
overshooting of the asymptotic value of the correlation
function produces a lowering of the cross section. In general the
results obtained with the gaussian correlation, which does not have
that overshooting, are larger than those obtained with the other
correlations. 

Also in this case this general behaviour strongly depends on the
kinematics.  We have shown in Fig. \ref{fig:sp}, that for a specific
case in superparallel back-to-back kinematics, the gaussian results
are the smaller ones. The study of this special case has shown that
the shapes of the angular distributions are strongly related to the
wave function of the correlated proton pair before being
emitted. Choosing special points of this angular distributions could
be misleading on the general behaviour of the cross sections.

Our study shows that the behaviour of the ($\gamma$,pp) cross sections
analyzed in terms of angular distribution of one of the emitted
protons, can be rather well interpreted in terms of SRC.  We have
shown that the cross sections follows a well controlled behaviour if,
after fixing the angle of one of the emitted protons, one analyzes the
complete angular distribution of the other proton. A selection of
specific kinematics, such as the symmetric or superparallel ones,
could provide results out of the systematics. The evident experimental
limitations with respect to a full 4 $\pi$ coverage of the angular
distribution, implies some specific kinematics to be selected.  It is
therefore necessary a joint work between experimentalists and
theoreticians to analyze the peculiarities of the chosen kinematics.

The above considerations can be applied also to the (e,e'pp)
reactions. There are, however, two features making the emission of two
protons induced by real photon a better tool for the study of the SRC.
First, in the photon case the longitudinal response is not present,
therefore the analysis of the angular distributions is simplified.
Second, the $\bqu$ dependence of the $\Delta$ currents is such that
their minimum contribution is at the photon point.

\section*{Acknowledgments}
We thank Carlotta Giusti for her interest in our work and the numerous
discussions. This work has been partially supported by the agreement
INFN-CICYT, by the DGI-FEDER (BFM2002-03218), by the Junta de
Andalucia (FQM 220) and by the MIUR through the PRIN {\sl Fisica del
nucleo atomico e dei sistemi a molticorpi}.

%\vspace*{1cm}
%

\end{document}